# Information Leakages in the Green Bond Market[1]

Darren Shannon[1*]

Jin Gong[1]

Barry Sheehan[1]

[1] Kemmy Business School, University of Limerick, Ireland

[*] Corresponding author: darren.shannon@ul.ie

[1] We are grateful for the comments provided by discussants at the 2023 World Finance Conference, the 2023 Conference on International, Sustainable and Climate Finance, and the 2025 Sustainable and Impact Investments International Conference. Finally, we would like to thank Dr. Francois Toerien for the provision of supplementary data that aided in the completion of this research study.

# Abstract

Public announcement dates are used in the green bond literature to measure equity market reactions to upcoming green bond issues. We find a sizeable number of green bond announcements were pre-dated by anonymous information leakages on the Bloomberg Terminal. From a candidate set of 2,036 'Bloomberg News' and 'Bloomberg First Word' headlines gathered between 2016 and 2022, we identify 259 instances of green bond-related information being released before being publicly announced by the issuing firm. These pre-announcement leaks significantly alter the equity trading dynamics of the issuing firms over intraday and daily event windows. Significant negative abnormal returns and increased trading volumes are observed following news leaks about upcoming green bond issues. These negative investor reactions are concentrated amongst financial firms, and leaks that arrive pre-market or early in market trading. We find equity price movements following news leaks can be explained to a greater degree than following public announcements. Sectoral differences are also observed in the key drivers behind investor reactions to green bond leaks by non-financials (Tobin's Q and free cash flow) and financials (ROA). Our results suggest that information leakages have a strong impact on market behaviour, and should be accounted for in green bond literature. Our findings also have broader ramifications for financial literature going forward. Privileged access to financially material information, courtesy of the ubiquitous use of Bloomberg Terminals by professional investors, highlights the need for event studies to consider wider sets of communication channels to confirm the date at which information first becomes available.

**Highlights**

- We find evidence of news leaks pre-dating public announcements of green bonds
- Leaks significantly influence trading dynamics in the underlying equity market
- Immediate and sustained impact: negative returns but increased trading volumes
- Timing of news and firm sector (financials vs. non-financials) influences reactions
- Perceived transformative impact of green bonds potential driver of market reactions

# 1 Introduction

2023 was the warmest year on record, with global temperatures climbing to 1.45°C above pre-industrial levels (World Meteorological Organization 2024). With economic losses of $280 billion observed in 2023 alone due to climate-related disasters (Swiss Re Institute 2024), financing plays a key role in preserving sensitive ecosystems and aiding in the recovery of economies within climate-affected regions. Sustainable financing mechanisms fulfil this role while providing a source of utility to investors seeking financial gain (Edmans and Kacperczyk 2022). Sustainable companies are likely to offer higher risk-adjusted returns if the market has not fully priced in the benefits that these business opportunities bring (ibid, p.2), making them attractive to investors seeking higher financial returns.

Green bonds[2] have emerged as a popular fixed-income solution to fulfil sustainable investing demands. Green bonds have resulted in a significant redirection of resources toward green-related initiatives and enhanced environmental performance[3] (Flammer 2020; Yeow and Ng 2021). Issuing firms enjoy increased levels of investor attention (Caramichael and Rapp 2022; Serafeim and Yoon 2022) and enhanced corporate performance (Tan *et al.* 2022; Dong *et al.* 2024). However, with the increasing popularity of Green bonds, there has been a parallel increase in the number of firms presenting misleading or exaggerated claims about the impact of their sustainability initiatives (de Freitas Netto *et al.* 2020), otherwise known as 'greenwashing'. This can not only lead to consumer mistrust but diminish the credibility of the companies that make genuine environmental efforts (Yang *et al.* 2020). Evidence of greenwashing concerns has previously been raised by Berrone

---

[2] Green bonds are a subset of financial products called 'Green, Social, and Sustainability Bonds (GSSBs)'; bonds where the proceeds are dedicated to the pursuit of environmental and social projects (ICMA 2023). We focus on Green Bonds as they carry an explicit commitment to fund projects with environmental benefits (p.3). Although the green bond market is relatively small relative to the global bond market, its rapid growth shows a strong interest from issuers and investors who support environmental sustainability (Maltais and Nykvist 2020).

[3] Both Flammer (2020) and Yeow and Ng (2021) note this only holds for bonds that are certified by independent third parties, who ascertain whether the bond proceeds will be used for 'credible' investments.

*et al.* (2017), Alessi *et al.* (2021), Flammer (2021), and Dong *et al.* (2024), and may lessen the positive impact and environmental performance observed amongst green bond issuers, on average (Bams *et al.* 2022). Secondly, there is a concern that the increased investor attention that arrives with the announcement of an upcoming green bond could incentivize unsuitable creditors to enter the green bond market.

This research study explores the confluence of these two concerns by setting two central objectives: to explore equity market reactions to green bond announcements, and to explore the potential for 'green' investing behaviours to be exploited by market insiders. We examine whether firms leverage positive investor sentiment toward green bonds and leak green bond-related news early. Information leakage refers to the dissemination of material, non-public, or privileged information through different communication channels (e.g. news media) to influence market behaviours before formal corporate announcements. Investors with access to non-public privileged information can benefit financially at the expense of uninformed investors (Gurgul and Majdosz 2007). Information leakages can significantly affect the trading volume and stock prices before an event is officially announced (Aitken and Czernkowski 1992), and can alter the dynamics of financial markets by changing how information is incorporated into prices (Brunnermeier 2005). Sletten (2012) outlines why firms may choose to leak such 'good news' early, with the chief motivation being managerial desires to maintain high stock prices for their firm.

Our study examines the intraday and daily trading activity of a set of stocks for which we find evidence of information leakage, in terms of equity price and trading volume changes. We define information leakage as news released about an upcoming green bond issue before the issue is publicly announced[4] by the issuing firm. We derive our unique dataset from news headlines extracted from the Bloomberg terminal over a 7-

---

[4] The public announcement date is when news of an upcoming green bond issue should theoretically be available to investors for the first time. This date has been used for measuring investor reactions to green bond announcements in seminal studies such as Tang and Zhang (2020) and Flammer (2021). We posit some investors had access to knowledge of upcoming green bond issuances before the public announcement date.

year period, spanning the years 2016-2022. To isolate 'leaks', we include conditional search terms to confirm i) the headlines relate to news about an upcoming green bond issue, and ii) the primary ticker involved in the upcoming announcement. Once candidate news headlines are gathered, we cross-check the timestamp of the headline against the public announcement date of all green bonds subsequently issued by the firm over the following three months, according to the Bloomberg database. A news headline is categorized as a 'leak' if it occurs at least one trading day before the public announcement date logged in these databases. From an initial sample of 2,036 candidate headlines, we find 231 instances of information leakages from 170 firms relating to upcoming green bond announcements. The timestamps of these news releases are used as placemarks to extract pricing and trading information about the primary equity ticker associated with the headline.

We examine trading volume patterns (in addition to equity price changes) as sharp increases in volume might suggest that some investors are acting on information they perceive as advantageous before it becomes fully reflected in the stock price (Foster and Viswanathan 1993). When combined with price data, changes in trading volume can help confirm the strength of price movements. We also prioritise the study of intraday as well as daily trading patterns. Efficient markets are assumed to absorb the implications of financial news on stock prices within minutes of dissemination (Francis *et al.* 1992), but factors such as press release timing (DellaVigna and Pollet 2009), stock liquidity (Chordia *et al.* 2009), and social network centrality (Hirshleifer *et al.* 2024) can affect price drift and volatility patterns. Thus, to ensure we thoroughly capture market reactions to green bond announcement leaks, we examine intraday trading patterns in the moments following the leak as well as the days following the leak.

Using an event study methodology, we find that green bond-related information leakages significantly influence trading volumes for the tickers mentioned in the news headlines. Price impacts are initially muted for the population sample of equities but demonstrate a significant negative trend one trading week after the leaks. These findings are confirmed when calculating abnormal returns[5] according to the market-adjusted model, the Capital Asset Pricing Model (CAPM), the Fama-French three-factor model, and the Carhart four-factor model[6]. While these population sample results suggest investors view green bond issuances as a negative signal about the issuing firm's future profitability or financial stability, rather than a positive commitment to sustainability, there is further nuance to our findings. We further reveal that negative investor reactions are strongly concentrated in leaks linked to financial firms, and leaks that arrive early in the market trading window (pre-market or within the first hour of trading). In contrast, investors tend to react positively to leaks by non-financial firms. For this subset of equities, returns are more predictable and are generally associated with more positive (yet non-significant) outcomes.

Our study furthers investment analysis and green bond literature by exploring the dynamics between these 'green' fixed-income announcements and equity price behaviours. A plethora of studies have shown that firms announcing upcoming green bond issues have enjoyed significant boosts to the price of their underlying equity, positively impacting firm value (Baulkaran 2019; Kuchin *et al.* 2019; Tang and Zhang 2020; Wang *et al.* 2020; Flammer 2021; Camacci 2022; Verma and Bansal 2023; Pirgaip *et al.* 2024). Furthermore, prior evidence of information leakage around green bond announcements has been found (e.g. Baulkaran (2019)), without the effect of the leaks being confirmed empirically. To the best of the authors' knowledge, our study is the first to explore the mechanisms by which these news leaks impact financial markets. We explore the factors driving the changes in trading patterns, and the industry classifications most likely to benefit from green bond-related information leaks. Our findings provide insights on equity market reactions resulting from differential levels of access to information, with the Bloomberg terminal primarily used by institutional investors over retail investors. Furthermore, we reveal equity reactions are more predictable in the days following information leaks than following public announcements, which warrants further analysis.

---

[5] Abnormal returns are defined as the difference between the performance of the underlying equity and the underlying market during a pre-defined event window. Day 0 of our event window is the leakage date of an upcoming green bond announcement.

[6] These nuanced event study models include additional variables to account for common drivers of change between equity and market prices. Including these models in our analysis allows us to further isolate the effect of the 'leak' itself, and thus confirm its impact on price and trading volumes.

The remainder of our study is structured as follows. Section 2 details the data gathered for our study, including a breakdown of the temporal distribution and intraday timing of the leaks, and the steps taken to reduce the set of green bond-related headlines down to identifiable leaks. Section 3 outlines the methodologies used to measure abnormal changes in equity prices and trading volumes, on both an intraday and daily level. Section 4 presents the findings of our analysis. Section 5 discusses the findings in the context of prior literature while exploring the implications of our findings. Section 6 concludes.

## 2   Data Collection and Description

### 2.1. Green Bonds

The green bond data used in this study is compiled from Bloomberg's fixed income database by extracting details relating to all active and matured bonds announced between May 22$^{nd}$, 2007 and January 29$^{th}$, 2023[7], where the use of proceeds was designated as 'Green Bond / Loan'. 7,375 green bonds or loans were announced during this period. Identifying information for the bond issuer is also extracted and indicates that a majority of the bonds are issued by private subsidiaries of larger corporations. Hence, we gather information relating to the ultimate parent, which allows us to identify each firm's status as a private or public company. Private corporations are removed from the analysis. We use the details associated with the remaining public parent companies to identify their ticker and the primary stock exchange on which they are listed. Ultimately, our approach returned 3,462 bonds listed by public companies, dating from 29th January 2013 to 30th January 2023. However, some companies announced multiple bond issues on the same day. It can be reasonably assumed that the trading activity observed during this trading day will reflect investors' reactions to the collective set of green bonds announced that day. Hence, we condense multiple same-day announcements into a single green bond 'announcement event'. There were 2,631 such 'announcement events' during the study period. Given the comprehensiveness of Bloomberg's fixed income database, our sample represents a significant coverage of the public green bond universe during this period. The distribution of green bonds announced on global exchanges during the study period is available in Figure 1.

---

[7] May 22$^{nd}$, 2007 is chosen as it is the date the world's first 'Green Bond' was issued (an €800m Climate Awareness Bond issued by the European Investment Bank), while January 30$^{th}$, 2023 is chosen as it marks 10 years since the first green bond issued by a publicly-traded company (a €3.6m Green Bond issued by Crédit Agricole).

## 2.2. Identifying Information Leakages

The primary objective of this study is to measure the impact of information leakages on firm value and trading activities, as measured by changes in the stock price and trading volumes in the moments and days following the leak. To identify the release of privileged information relating to upcoming green bond announcements before being publicly announced, we perform a search on the 'Bloomberg News' and 'Bloomberg First Word' data screens. The data screens often contain anonymous 'tips' provided to Bloomberg about upcoming financial operations, which are subsequently broadcast to the market. Many of these headlines have associated tickers attached. As such, information that is as-of-yet unknown to the wider public is broadcast to Bloomberg Professional subscribers, along with the companies involved and their associated IDs.

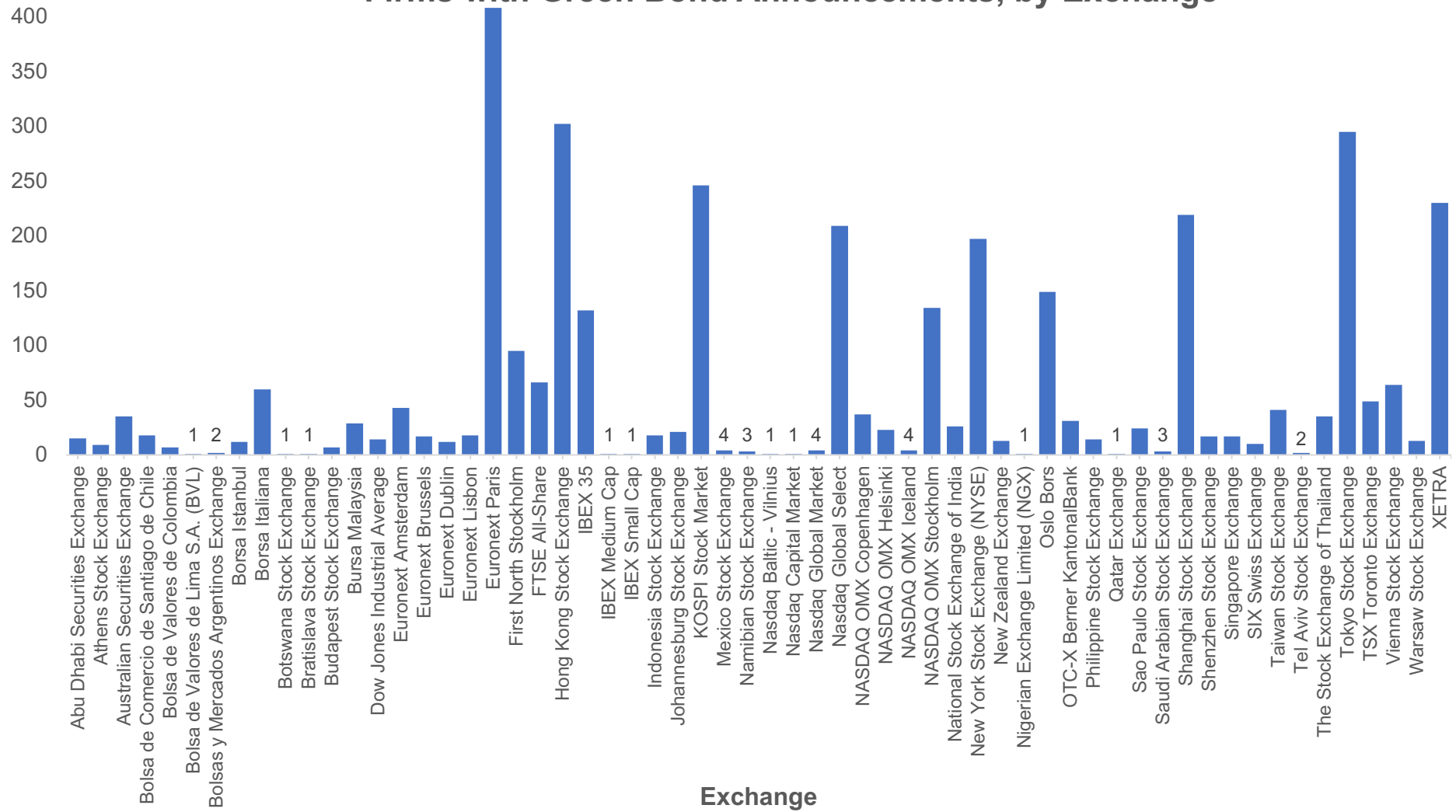

**Figure 1:** Distribution of green bonds announced by publicly traded firms during the period January 2013- January 2023, by the primary stock exchange on which the issuing firm is listed.

We include search operators to isolate news items to Green Bonds. Each candidate headline must contain "Mandate" and "Green", and the associated article must be assigned a 'Green Bond' label by Bloomberg. The "Mandate" search operator is included as a bond mandate letter is likely to be the first document encountered by a bond issuer when starting the process of preparing for a bond issue. Thus, it is likely the first opportunity for information about an upcoming bond issue to be leaked. Sample search results for Bloomberg News and Bloomberg First Word when including these search operators are available in Figure 2, along with a sample article related to one of the extracted headlines.

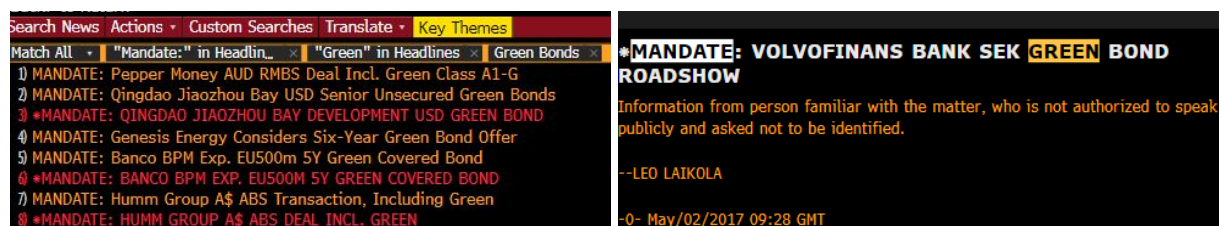

**Figure 2:** Bloomberg screens reporting (left) news relating to upcoming green bond announcements and discussions after including "Mandate", "Green", and 'Green Bond classification' search operators, and (right) the news article for one of the extracted headlines providing leaked information about an upcoming green bond announcement.

Our search yielded 2,036 headline results that reduced to 1,197 after removing duplicate headlines and multiple 'update' articles related to the same green bond issue. Duplicate headlines largely related to the same headline being posted to both Bloomberg News and Bloomberg First Word. Of these, the news headline with more information included in the article was retained. There were also nine instances of lengthy gaps between news article 'updates' relating to the same green bond issue. The earliest known headline for each green bond issue was retained, and newer headlines were removed from the analysis. Of the remaining 1,197 news headlines, 1,050 mentioned primary tickers.

Once the set of unique headlines including primary tickers were identified, we cross-checked the associated equities against the database of 2,631 green bond announcement 'events' derived in §2.1. If the primary ticker associated with the news article publicly announced an upcoming green bond issue within two months of the news headline, and if the timestamp of the news headline occurred at least one business day before the public announcement date of the green bond announcement 'event', the news headline is classified as a 'leak'. Of the 1,050 headlines with tickers, 259 were publicly traded tickers who subsequently announced upcoming green bond issues in the 2 months following the publication of the leaked news headline. 60% of the leaks occurred on Monday and Tuesday (Figure 3, left). There was a sharp uptick in the number of leaks released in 2021, accounting for 40% of leaks in our sample (Figure 3, right). When examining the timestamps of the leaked headlines against the opening hours of the exchange on which the primary ticker is listed, it becomes apparent that leaked information is generally released toward the beginning of the trading day, with the notable exception of the Japanese exchange (Figure 4).

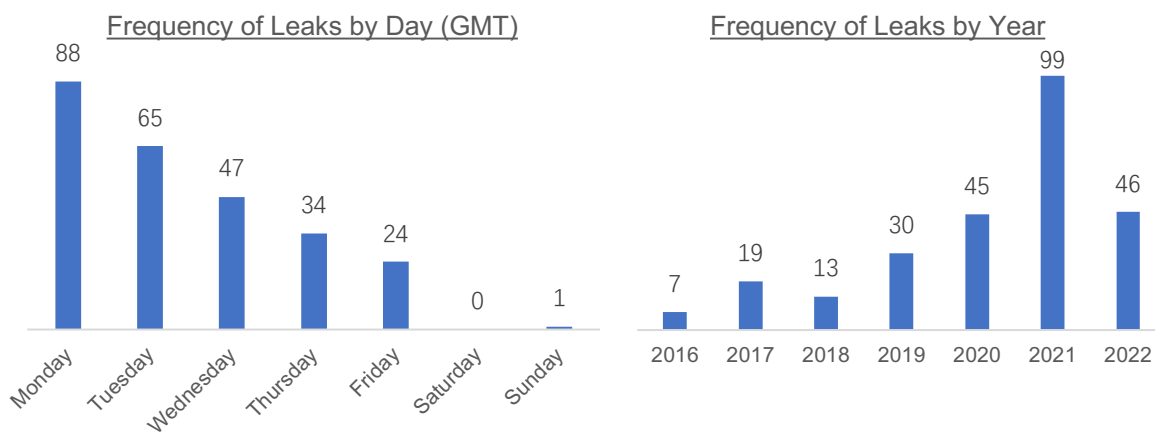

**Figure 3:** Temporal distribution of 259 green bond announcement leaks, according to (left) day of release and (right) year of release.

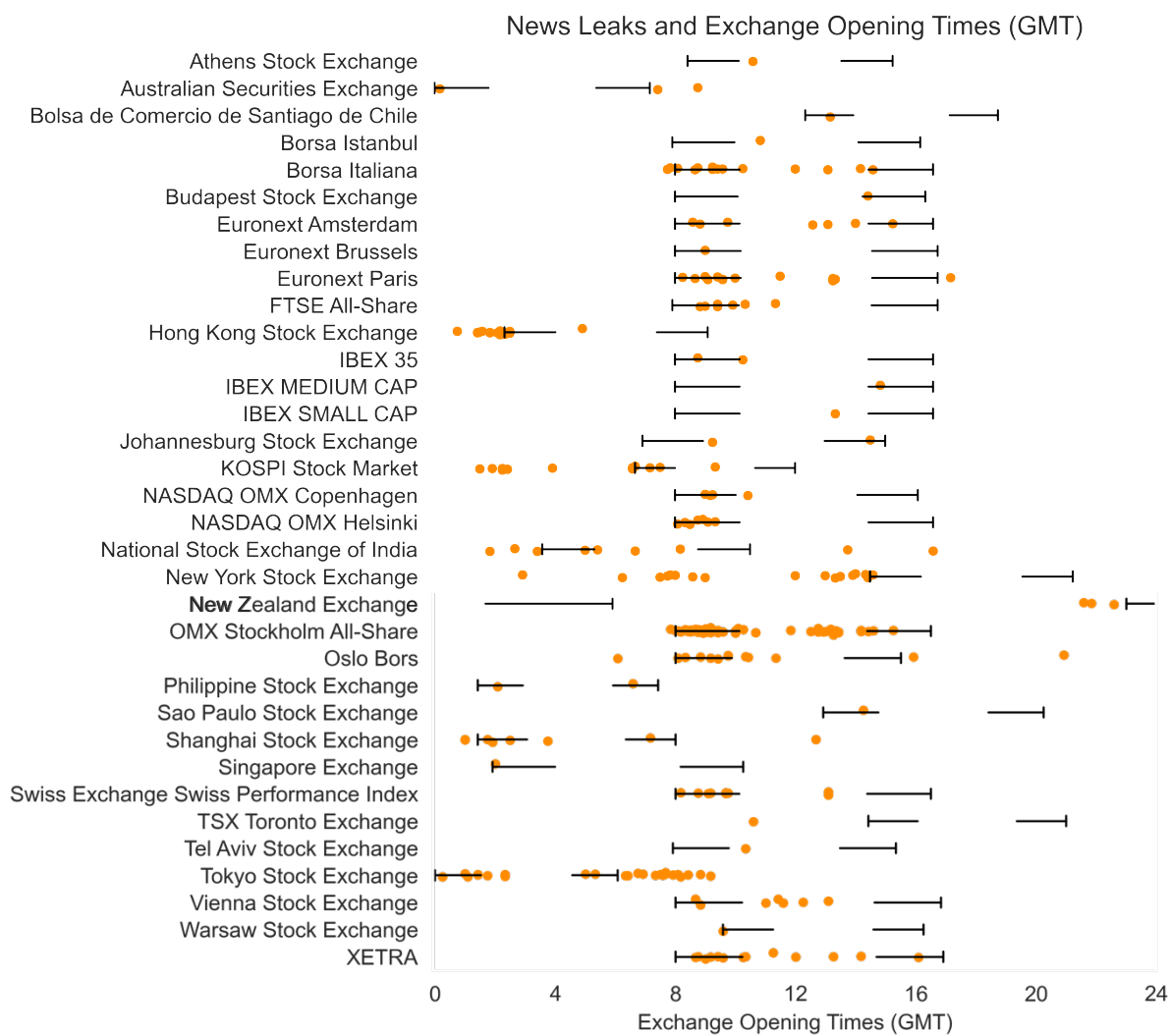

**Figure 4:** Distribution of timings of leaked headlines (in orange) relative to the opening hours of the exchange on which the primary ticker is listed.

## 2.3. Supplementary Information and Data Descriptions

To support our insights, we gather supplementary information that could influence the reactions of reasonably informed investors to leaked news of upcoming green bonds. We limit the information we gather to data available to investors at the news event, ruling out the collection of influential information such as yield spreads, credit ratings, and green bond certification status. We collect a range of factors relating to the planned issue, such as the bond's currency, maturity, coupon, and the issue date and size. Since there are many instances where multiple green bonds were announced by a company on the same day, for each Green Bond 'event' we collect information on i) the number of bonds that were announced, ii) the cumulative size of the anticipated issuance (in USD terms), iii) the average coupon paid by the collective issue, and iv) the presence of an option embedded in the payout structure.

We also collect information relating to the firm's financial fundamentals over the three years before the announcement of the bond and their business classifications to determine the impact of the firm's financial stability and performance on market reactions to leaked news. We account for firm size (total assets), profitability (return on assets), leverage (debt-to-equity), liquidity (free cash flow), experience with lending in the green bond market (first-time green bond issuer Y/N), and measures of market perception (market capitalisation and Tobin's Q). We use the averaged three-year financial data for our analysis, as suggested by Levis (1993), as it smooths out annual fluctuations and reduces the impact of outliers. If the green bond announcement arrived in the first 45 days of the year[8], we instead use trailing 12-month figures correct as of the latest filing that was available to investors at the time of the announcement (typically Q3 results). To maintain data reliability and accessibility standards relating to a firm's financial strength, we remove from our sample 3 companies that went public in the three years before the green bond leak, reducing our eligible sample to 256.

**Table 1:** Descriptive statistics of the 256 green bond issues linked to leaks. Firm-level characteristics are averaged over the three years preceding the announcement event. Variable descriptions are available in Appendix A.

| Panel A: Green Bond Characteristics | N | Min | Max | Mean | Std. dev. |
|---|---|---|---|---|---|
| Issue Size | 256 | 3.72 | 4197.28 | 534.49 | 566.65 |
| Term | 256 | 2 | 51 | 7.17 | 5.11 |
| Cpn | 256 | 0 | 8.50 | 2.09 | 1.84 |
| Option in Issue | 256 | – | – | 0.40 | – |
| Bonds in Issue | 256 | 1 | 4 | 1.36 | 0.69 |
| **Panel B: Firm-level Characteristics** | **N** | **Min** | **Max** | **Mean** | **Std. dev.** |
| FTI | 256 | - | - | 0.51 | - |
| MktCap | 256 | 90.82 | 245,680 | 21,904 | 39,389 |
| Assets | 256 | 13.75 | 3,738,164 | 266,471 | 654,514 |
| ROA | 256 | -4.77 | 20.63 | 3.24 | 3.40 |
| D/E | 256 | 0 | 11.11 | 2.13 | 2.24 |
| FCF | 256 | -19,276 | 72,206 | 2,940 | 10,567 |
| Tobin's Q | 256 | 0.01 | 6.61 | 0.58 | 0.82 |
| **Panel C: Outcome Variables** | **N** | **Min** | **Max** | **Mean** | **Std. dev.** |
| CAR[0,1] | 214 | -7.30 | 9.88 | -0.15 | 2.50 |
| CAR[0,2] | 214 | -9.46 | 9.68 | -0.36 | 2.86 |

---

[8] during this period the latest annual filings would not have been made available to public investors

To allow for comparability amongst issuers of different sizes and markets, all relevant monetary information is rebased to US dollars using the foreign exchange rate correct at the time of the news leak. Table 1 summarises the descriptive statistics of the financial information used in our regression analysis in Section 4. Appendix A and Appendix B detail the definition of the variables and the correlation matrix for the variables, respectively. The average announced issue size of these anticipated green bonds was $534.5 million and carried an average time to maturity of 7.17 years[9], while the average coupon rate on offer was 2.09% (Table 1).

## 2.4. Event Study Data

The event study methodology we employ in our study examines the stock market reaction around both the leakage date and the announcement date of the green bond issuance. This impact is evaluated by calculating the abnormal returns and abnormal trading volumes observed on both the leakage date and announcement dates for each of the involved companies. We calculate our baseline and intraday abnormal returns according to the market-adjusted model (Brown and Warner 1985). In the market-adjusted model, abnormal returns are calculated as the difference between a firm's stock return and the return observed in the market on which the stock is listed. To ensure the robustness of our analysis, we further compute abnormal returns according to the Fama and French (1993) 3-factor model and Carhart (1997) 4-factor model, along with the CAPM. We compute the abnormal returns for these models using market loading factors drawn from the CRSP database (French 2023), supplemented with further hand-collected market loading factors for the stock exchanges based in South Africa, China, and India. The market loading factors allow for regression analyses on the underlying drivers of abnormal returns during periods of information leakage and allow us to further isolate the effect of the news leak. The market loading factors, along with their periodicity, are available in Appendix C.

---

[9] 6 of the 256 green bond leak events included only perpetual bonds with a theoretical time to maturity of $\infty$. The time to maturity for these bonds were replaced with their Macaulay Duration (calculated as $\frac{1+yield}{yield}$), as this represents the weighted average term to maturity of the cash flows from a bond. Assuming a duration gap of zero, this term is posited as a proxy for the view a reasonably informed investor may take on the life of a perpetual bond.

Issues emerge when calculating abnormal returns on thinly traded stocks, particularly when there are few events available for each market being examined (Bartholdy *et al.* 2007). Our study is global in scope and the events we are considering are dispersed across many markets. As such, the inclusion of thinly-traded equities may disproportionately skew the statistical estimates we generate in the Fama-French and Carhart regression models (Sercu *et al.* 2008). To ameliorate this issue, we impose two liquidity conditions: multiple trade orders must be placed in the hour following the leak (or within the first hour of trading if the leak occurred pre-market), and the 24-hour trading volume following the leak must be greater than 10,000 units. From the remaining 256 tickers, 18 did not meet these liquidity conditions and so are removed from the analysis. Of the remaining 238 tickers, market loading factors were available for 214 (a coverage rate of 90%). The 214 events stem from 159 companies across 23 countries.

## 3 Methodology

### 3.1. Abnormal Returns

#### 3.1.1. Market-adjusted Model

The Fama and French (1993) three-factor model and the Carhart (1997) four-factor model are used alongside the market-adjusted model and the CAPM to allow for the inclusion of additional non-market factors. Country-specific stock market indices are used in our research to calculate market returns. Abnormal returns $AR_{i,t}$ in the market-adjusted return model are determined by the formula in Equation 1:

$$AR_{i,t} = R_{i,t} - R_{m,t} \qquad (1)$$

Where $R_{m,t}$ is the return observed on the underlying market upon which security $i$ is listed on day $t$. The market-adjusted model assumes that under conventional trading conditions, the expected return on equity should, on average, be equal to the expected return on an equivalent investment in the underlying market, i.e. $E(R_i) = E(R_m) \, \forall \, i$ (Strong 1992). Any significant deviations that are observed, on average, between security and market returns over the relevant event window are indicative of the event of interest having a notable influence on the security's equity price.

In line with the event study methodology of Fama *et al.* (1969), we calculate individual abnormal returns (ARs) in line with the market-adjusted return methodology. We thereafter aggregate the ARs for each day in the event window to calculate the average abnormal returns (AARs) observed during the event window where, given $N$ ARs on day $t$ in the event window, $AAR_t = \frac{1}{N}\sum_{i=1,\ldots,N} AR_{i,t}$. We also calculate a rolling sum of the AARs to form cumulative average abnormal returns (CAARs). The choice of event windows is a critical component when conducting event study analyses. We base our event windows on those used in prior research on green bond issuance events (Michaelides *et al.* 2015; Lebelle *et al.* 2020; Laborda and Sánchez-Guerra 2021; Camacci 2022). Both AARs and CAARs allow us to gain insights into the instantaneous and daily reactions of equity investors to the green bond leaks.

### *3.1.2. Fama-French 3-factor Model and Carhart 4-factor Model*

The CAPM, developed by Sharpe (1964) and Lintner (1965) and stated in Equation 2, accounts for the differential levels of risk associated with an equity (idiosyncratic risk) and the market on which it is listed (systematic risk). They argued that the expected returns on individual stocks are proportional to the stock's level of systematic risk (beta), and described the relationship between the expected return of an asset and its specific exposure to market risk, or beta, by the formula:

$$R_i = R_f + \beta(R_M - R_f) \qquad (2)$$

Where $R_i$ is the (expected) return on asset $i$, $R_f$ is the risk-free rate, $\beta$ is the beta of the asset and $R_M$ is return on the market. Abnormal returns can be then calculated by comparing observed returns during the event window to the expected returns predicted by the parameters derived from the pre-event estimation window.

Fama and French (1993) added two firm-specific factors in addition to the beta factor proposed in the CAPM; the book-to-market value of a company ('value'), and the market capitalisation of the firm ('size'). Carhart (1997) further extended Fama-French's three-factor model to account for the rate of change in market price movements, also called the momentum factor (MOM). This factor was introduced to capture the expectation that market leaders tend to sustain larger positive movements than market underperformers. The MOM factor is calculated as the daily return of the equal-weighted average of the highest-performing stocks (winners) minus the daily return of the equal-weighted average of the lowest-performing stocks (losers). The FF3M is expressed in equation 3, while the Carhart Four Factor model is expressed in equation 4:

$$R_i = R_f + \beta_1(R_M - R_f) + \beta_2(SMB) + \beta_3(HML) \qquad (3)$$

$$R_i = R_f + \beta_1(R_M - R_f) + \beta_2(SMB) + \beta_3(HML) + \beta_4(MOM) \qquad (4)$$

Where $R_i$ is the (expected) return on asset $i$, $R_f$ is the risk-free rate, $R_M$ is the return on the market, $\beta_1$ is the sensitivity of the asset to the market, $\beta_2$ (SMB) is the historic excess returns of small-cap companies over large-cap companies, $\beta_3$ (HML) is the historic excess return of value stocks (high book-to-market ratio) over growth (low book-to-market ratio), and $\beta_4$ (MOM) is the historic excess return of market winners over market underperformers.

For the aforementioned models, abnormal returns are calculated by comparing observed returns to the model-predicted (expected) returns formed from observations from the estimation window. In the case of setting the announcement date as the event date (day 0), the time horizon in the baseline event window ranges from 10 days before and after the event date [-10, 10] with the estimation window being [-310, -11]. When using the leakage date as the event date (day 0), the baseline event window is set from day 0 to day 10 with the estimation window spanning from day -300 to day -1. Additionally, the time intervals [11, 21] and [21, 30] are examined to identify any notable patterns following the event window (Flammer 2021).

## 3.2. Abnormal Trading Volumes

While abnormal returns capture how the market prices new information, abnormal trading volume indicates how actively investors are reacting to that information. High abnormal trading volume suggests heightened investor interest and potential information asymmetry (Chae 2005; Lei and Wang 2014). We calculate abnormal trading volumes in line with Jansen (2015), as the natural log of shares traded during the event window (scaled by shares outstanding) relative to the average trading volume during the estimation window (Equation 5):

$$\text{Log Turnover } (\tau_{i,t}) = \ln\left(\frac{\text{Trading Volume}_{i,t}}{\text{Shares Outstanding}_{i,t}}\right) \quad (5)$$

With abnormal trading volume computed as $\xi_{i,t} = \tau_{i,t} - \bar{\tau}_i$, where $\bar{\tau}_t$ is the baseline expected trading volume calculated by measuring the average log turnover of a pre-event estimation window:

$$\bar{\tau}_t = \frac{\sum_{t=-T}^{t=-1} \tau_{i,t}}{T}$$

When calculating intraday abnormal volumes, we set our estimation window to be $T = 96$. The abnormal volume for each 5-minute event window following a leak is calculated by comparing the trading volume to the average 5-minute trading volume observed in the 8 trading hours immediately preceding the leak[10]. When calculating daily abnormal volumes, we set our estimation window to be $T = 20$, meaning daily abnormal trading volumes following a leak are calculated by comparing them to the average daily trading volume observed in the month before the leak. We calculate average abnormal volumes (AAVs) for the intraday windows and additionally calculate cumulative average abnormal volumes (CAAVs) when widening the event window to include daily changes in trading volumes.

---

[10] We do scale by shares outstanding when examining intraday trading volumes, as variations in the number of shares outstanding are minimal over a single trading day. This reduces the log turnover equation to $\tau_{i,t} = \ln(\text{Trading Volume}_{i,t})$.

# 4   Results and Discussion

## 4.1. Trading Dynamics following Information Leakages

### 4.1.1. Intraday Event Window: Abnormal Returns and Trading Volumes

Our results initially reveal a muted price impact following the leak of upcoming green bond news. Table 2 describes the intraday change in the underlying equity price following the release of the news headline, measured from minute 0 to hour 8 (a full trading day). The figures in Table 2 represent market-adjusted returns, i.e. equity returns in excess of the underlying market. Our intraday perspective means the number of equities qualifying for analysis decreases as the time elapsed following the news leak increases. Initial reactions are negative when examining the full sample of equities (Table 2, Panel A), with mixed reactions observed following the hour mark of the news leakage. None of these equity price changes are significant.

Significant variations in equity price changes are observed when accounting for the timing of the news leakage (Table 2, Panel B). We split our sample into Early Market News ($n$ = 120) and Late Market News ($n$ = 118). Early Market News is classified as information released either pre-market (including information released after the previous day's market close) or within the first hour of trading. Late Market News is classified as information released subsequent to one hour after market open. Pre-market and early-market news tend to induce significant negative reactions, with an average fall of 10-12 basis points observed in the 30 trading minutes following the leak. These losses in share price intensify as the trading day continues, with losses of 17-19 basis points observed 2-5 trading hours after the leak. Late market news, in contrast, induces positive reactions. On average, equity prices increase by 10-13 basis points in the two hours following the leak. These positive investor reactions are sustained throughout the remainder of the trading day, but their significance wanes as the variance of observed investor reactions across the whole sample increases. We surmise the differential price impacts may be attributable to timing strategies employed by management teams tasked with leaking green bond-related news. Patell and Wolfson (1982) empirically confirm the conventional 'market wisdom' that 'good news' is released during trading hours, while 'bad news' is withheld until after market close. They found less favourable news is more frequently released outside of active trading periods, while favourable news is more likely to be released during active trading periods. We observe the same phenomenon in Table 2, Panel B.

**Table 2:** Cumulative Average Abnormal Returns (CAARs) observed in the 8 trading hours following the release of private information relating to an upcoming green bond announcement. Price impacts are only considered during active trading windows, resulting in a reduction of qualifying equities as the time elapsed increases. *, **, and *** indicate significance at the 10%, 5%, and 1% level, respectively.

| Time from Leaked News | Panel A. Total Sample | | Panel B. Timing of Leak | | | | Panel C. Source of Leak | | | |
|---|---|---|---|---|---|---|---|---|---|---|
| | All | | Early Market News | | Late Market News | | Financial Institutions | | Non-financial Institutions | |
| | n | bp$^{sig}$ (std. err.) | n | bp$^{sig}$ (std. err.) | n | bp$^{sig}$ (std. err.) | n | bp$^{sig}$ (std. err.) | n | bp$^{sig}$ (std. err.) |
| 0 (mins) | 238 | – | 120 | – | 118 | – | 114 | – | 124 | – |
| 5 | 238 | -1.45 (1.72) | 120 | -3.39 (3.14) | 118 | +0.52 (1.35) | 114 | **-4.19* (2.28)** | 124 | +1.06 (2.54) |
| 10 | 238 | -2.18 (2.75) | 120 | -7.36 (5.06) | 118 | +3.10 (2.01) | 114 | -3.53 (2.52) | 124 | -0.93 (4.76) |
| 15 | 238 | -3.80 (2.84) | 120 | **-10.77** (5.16)** | 118 | +3.29 (2.13) | 114 | -3.18 (2.85) | 124 | -4.37 (4.79) |
| 20 | 238 | -2.66 (3.34) | 120 | -8.83 (6.06) | 118 | +3.63 (2.61) | 114 | -3.13 (3.45) | 124 | -2.22 (5.58) |
| 25 | 238 | -3.97 (4.05) | 120 | **-12.19* (7.26)** | 118 | +4.39 (3.36) | 114 | -5.64 (5.36) | 124 | -2.43 (6.03) |
| 30 | 238 | -3.39 (4.33) | 120 | -11.82 (7.54) | 118 | +5.18 (4.08) | 114 | -6.64 (5.44) | 124 | -0.41 (6.65) |
| 35 | 236 | -0.87 (4.68) | 120 | -8.02 (8.04) | 116 | +6.53 (4.56) | 114 | -5.07 (5.63) | 122 | +3.05 (7.36) |
| 40 | 236 | -2.16 (4.84) | 120 | -11.52 (8.28) | 116 | +7.52 (4.74) | 114 | -6.89 (5.84) | 122 | +2.26 (7.61) |
| 45 | 236 | -0.21 (4.99) | 120 | -10.44 (8.31) | 116 | **+10.37* (5.24)** | 114 | -4.09 (5.81) | 122 | +3.41 (7.98) |
| 50 | 235 | +2.19 (4.54) | 120 | -7.01 (7.26) | 115 | **+11.79** (5.23)** | 114 | +0.29 (4.86) | 121 | +3.98 (7.55) |
| 55 | 234 | +1.94 (4.79) | 120 | -8.87 (7.56) | 114 | **+13.31** (5.60)** | 114 | +0.02 (5.16) | 120 | +3.75 (7.97) |
| 1 (hours) | 234 | +1.68 (4.98) | 120 | -8.89 (7.89) | 114 | **+12.80** (5.82)** | 114 | -0.66 (5.19) | 120 | +3.90 (8.39) |
| 2 | 226 | -4.40 (5.59) | 120 | **-17.21* (8.99)** | 106 | **+10.11* (5.97)** | 111 | -2.36 (6.84) | 115 | -6.37 (8.82) |
| 3 | 217 | -7.19 (6.55) | 120 | **-18.38* (10.01)** | 97 | +6.66 (7.67) | 103 | -10.28 (8.51) | 114 | -4.39 (9.85) |
| 4 | 196 | -10.38 (7.28) | 119 | **-18.37* (9.69)** | 77 | +1.96 (10.86) | 90 | -16.75 (10.14) | 106 | -4.97 (10.37) |
| 5 | 184 | -7.43 (8.11) | 119 | **-19.05* (10.33)** | 65 | +13.84 (12.70) | 80 | **-19.77* (11.42)** | 104 | +2.06 (11.31) |
| 6 | 146 | -2.63 (9.42) | 102 | -5.50 (12.3) | 44 | +4.04 (12.93) | 63 | -21.33 (13.51) | 83 | +11.57 (12.86) |
| 7 | 84 | -7.06 (13.94) | 60 | -15.52 (17.79) | 24 | +14.09 (19.91) | 40 | -30.13 (19.46) | 44 | +13.92 (19.55) |
| 8 | 23 | +9.36 (24.11) | 23 | +9.36 (24.11) | 0 | – | 11 | -25.77 (37.44) | 12 | +41.57 (29.38) |

We also note a split in equity price changes based on the category of firms leaking information (Table 2, Panel C). Investor reactions tend to be significantly negative when the leaked news is attributed to financial firms. We classify financial firms as banks, government agencies, real estate companies, wider financial services firms, and those operating in consumer and commercial finance. There is an immediate fall of 4 basis points among these firms in the minutes following the leak, intensifying to a fall of 20 basis points after 5 hours of trading. These effects are significant at the 10% level. For the bulk of the trading day, slightly negative but non-significant equity returns are observed following information leakages from financial institutions. In contrast, non-financial firms see a rise in their share price in the hours following a news leak, albeit a non-significant increase. These mixed findings are not surprising, as heterogenous yet economically insignificant equity returns are often observed following bond issues in different markets and contract structures (see e.g. Grossmann and Ngo (2025) for an overview of recent and foundational literature).

The directionality of our findings, however, stands in contrast to prior empirical findings. Li *et al.* (2016) find financial institutions outperform non-financial institutions following the announcement of convertible bonds, while Duca *et al.* (2012) find non-financials experience negative abnormal stock returns. This is attributed to information asymmetry and adverse selection theory. When debt financing is announced as an alternative to the use of internal financing (particularly when equity-like convertible bond structures are announced), investors require a discount based on the risk-seeking financing decision (Myers and Majluf 1984; Miller and Rock 1985). However, our sample encompasses a unique macroeconomic environment that could plausibly suspend these established dynamics. Much of our sample is drawn from COVID-19 (56% of the leaks were recorded in 2020-2021) when debt financing was encouraged to spur economic activity. In addition, the 'green' label associated with the bond announcement ameliorates the asymmetry of information as to the planned proceeds of the financing and the strategy of the firm (Schmittmann and Gao 2022)[11].

Furthermore, Li *et al.* (2016) explain financial institutions outperform non-financials following convertible bond announcements as stringent regulations associated with the intended use of bond proceeds by financial institutions mitigate the information asymmetry between issuing firms and investors. The additional disclosure requirements required by financial institutions limit their incentives to take on excessive risks. However, these same regulations may hinder the capacity of financial institutions to innovate and develop transformative business models using green financing, ultimately acting as a negative signal for investors. We argue this is a plausible explanation for the non-standard abnormal returns observed in our sample.

---

[11] Further information asymmetry concerns are raised by Schmittmann and Gao (2022) on brown firms' ability to greenwash by issuing green bonds as a potentially misleading signal of their green credentials, but the market signal being sent remains clear.

Table 3 describes the intraday changes in trading volumes for the underlying equity price following the release of the news headline, measured from minute 0 to hour 8 (a full trading day). The relative changes indicate average abnormal returns for individual 5-minute periods following information leakages. Panel A highlights there is a significant increase in trading volumes when measured across the entire sample of leaks ($n$ = 238). In the minutes following an information leak, there is an 82% increase in trading volumes. Differences in trading dynamics are further highlighted when considering the timing of the leak (Table 3, Panel B) and the source of the leak (Table 3, Panel C). Early market news tends to produce significant increases in trading volumes in the 15 minutes following the leak (+25% to +161.5% rise, on average), while late market news is associated with a significant drop in trading volumes over the same period (-27% to -41% fall, on average).

**Table 3:** Average Abnormal Volumes (AAVs) observed in the 8 trading hours following the release of private information relating to an upcoming green bond announcement. Trading volume dynamics are only considered during active trading windows, resulting in a reduction of qualifying equities as the time elapsed increases. *, **, and *** indicate significance at the 10%, 5%, and 1% level, respectively.

| Time from Leaked News | Panel A. Total Sample | | Panel B. Timing of Leak | | | | Panel C. Source of Leak | | | |
|---|---|---|---|---|---|---|---|---|---|---|
| | All | | Early Market News | | Late Market News | | Financial Institutions | | Non-financial Institutions | |
| | $n$ | per$^{sig}$ (std. err.) | $n$ | per$^{sig}$ (std. err.) | $n$ | per$^{sig}$ (std. err.) | $n$ | per$^{sig}$ (std. err.) | $n$ | per$^{sig}$ (std. err.) |
| 0 (mins) | 238 | **+82.19** (33.74)** | 120 | **+161.52*** (58.08)** | 118 | +1.51 (32.45) | 114 | +38.75 (24.38) | 124 | **+122.12** (60.68)** |
| 5 | 238 | -1.26 (8.56) | 120 | **+24.88* (14.84)** | 118 | **-27.84*** (7.73)** | 114 | -14.66 (14.28) | 124 | +11.07 (9.82) |
| 10 | 238 | +57.01 (49.49) | 120 | +139.64 (97.34) | 118 | **-27.02*** (9.26)** | 114 | +97.55 (101.77) | 124 | +19.74 (16.91) |
| 15 | 238 | +5.76 (11.05) | 120 | **+51.97** (20.31)** | 118 | **-41.23*** (5.89)** | 114 | +4.92 (18.93) | 124 | +6.53 (12.19) |
| 20 | 238 | +12.04 (14.67) | 120 | +19.81 (13.05) | 118 | +4.15 (26.50) | 114 | -1.53 (14.02) | 124 | +24.52 (25.05) |
| 25 | 238 | -11.47 (7.48) | 120 | +8.21 (12.48) | 118 | **-31.49*** (7.80)** | 114 | -11.21 (12.15) | 124 | -11.71 (9.08) |
| 30 | 238 | +0.67 (14.56) | 120 | **-14.78* (8.07)** | 118 | +16.39 (28.19) | 114 | **-35.16*** (7.85)** | 124 | +33.61 (26.71) |
| 35 | 236 | -11.46 (8.41) | 120 | +4.56 (11.62) | 116 | **-28.03** (12.05)** | 114 | **-20.15** (9.54)** | 122 | -3.33 (13.62) |
| 40 | 236 | -11.63 (9.51) | 120 | -3.41 (14.54) | 116 | -20.14 (12.19) | 114 | **-23.87** (9.14)** | 122 | -0.21 (16.27) |

| 45 | 236 | **-27.19*** (6.56)** | 120 | -15.53 (9.92) | 116 | **-39.25*** (8.43)** | 114 | **-27.52*** (9.92)** | 122 | **-26.88*** (8.71)** |
| 50 | 235 | **-25.27*** (6.15)** | 120 | -10.15 (9.59) | 115 | **-41.05*** (7.36)** | 114 | **-20.01** (10.00)** | 121 | **-30.23*** (7.35)** |
| 55 | 234 | -9.00 (19.31) | 120 | **-20.36** (8.49)** | 114 | +2.96 (38.68) | 114 | **-28.99*** (8.57)** | 120 | +9.99 (36.76) |
| 1 (hours) | 234 | **-20.07** (8.59)** | 120 | +1.88 (15.02) | 114 | **-43.19*** (7.27)** | 114 | -9.09 (13.15) | 120 | **-30.51*** (11.12)** |
| 2 | 226 | -13.16 (16.20) | 120 | **-35.38*** (6.4)** | 106 | +12.01 (33.69) | 111 | +9.25 (32.29) | 115 | **-34.78*** (6.16)** |
| 3 | 217 | -15.59 (10.17) | 120 | **-33.37*** (9.11)** | 97 | +6.42 (19.60) | 103 | +5.11 (19.57) | 114 | **-34.29*** (7.57)** |
| 4 | 196 | +28.07 (41.36) | 119 | -12.36 (24.00) | 77 | +90.54 (98.52) | 90 | +58.69 (84.46) | 106 | +2.07 (26.96) |
| 5 | 184 | **+107.89*** (34.52)** | 119 | **+134.66*** (44.20)** | 65 | +58.86 (54.73) | 80 | +52.17 (44.95) | 104 | **+150.75*** (50.14)** |
| 6 | 146 | +34.78 (29.34) | 102 | +8.87 (20.81) | 44 | +94.84 (84.59) | 63 | +35.08 (33.17) | 83 | +34.55 (45.24) |
| 7 | 85 | +113.28 (80.78) | 61 | **-30.64*** (10.09)** | 24 | **+479.06* (274.98)** | 40 | +192.95 (160.79) | 45 | +42.46 (54.00) |
| 8 | 23 | +465.91 (326.57) | 23 | +465.91 (326.57) | 0 | – | 11 | +554.69 (509.25) | 12 | +384.53 (437.54) |

### *4.1.2. Daily Event Window: Abnormal Returns and Trading Volumes*

We next extend the event window to consider daily returns and trading volumes. Daily returns are calculated according to the CAPM, Fama-French 3-factor, and Carhart 4-factor models. Considering the global scope of our study, achieving total market factor coverage for the implementation of these models is a challenge. We attained market factors for 90% of our sample (Appendix C), reducing the number of equities eligible for further analysis from 238 to 214. Table 4 describes the CAPM-calculated average abnormal returns (AARs) observed after news is leaked about an upcoming green bond announcement (Panel B) and following the public announcement of an upcoming green bond issue (Panel A). An immediate and significant negative reaction to public announcements of green bonds is observed, with a fall of 20 and 12 basis points over the ensuing two trading days (Panel A). There is a delayed response to the leaked news (Panel B), yet the same downward-trending pattern is present. The two trading days following the release of the information saw modest yet non-significant falls of 5 and 10 basis points in the share price of the associated firm. On the third trading day following the news leakage, however, there is a 21 basis point drop in the underlying share price on average (Panel B). Minimal price impacts are otherwise observed over the [0,10] trading day event window[12].

---

[12] We also calculated abnormal returns for the 10 trading days leading up to the public announcement of the green bond issue, shifting our estimation window correspondingly to maintain accuracy. There are significantly positive price changes observed over the [-8, -7] trading day window pre-public announcement. No further significant price changes are observed.

**Table 4:** Average Abnormal Returns (AARs) observed over the 10 trading days following the release of green bond-related information. The abnormal returns are calculated using the Capital Asset Pricing Model (CAPM). *, **, and *** indicate significance at the 10%, 5%, and 1% level, respectively.

| Day from Announcement | n | Panel A. Public Announcement | | | Panel B. Leaked News | | |
|---|---|---|---|---|---|---|---|
| | | AAR$^{sig}$ | t-stat | p-value | AAR$^{sig}$ | t-stat | p-value |
| 0 | 214 | **-0.20%**** | **-2.142** | **0.016** | -0.05% | -0.411 | 0.341 |
| 1 | 214 | **-0.12%*** | **-1.377** | **0.085** | -0.10% | -0.936 | 0.175 |
| 2 | 214 | +0.03% | 0.295 | 0.384 | **-0.21%**** | **-2.218** | **0.014** |
| 3 | 214 | -0.05% | -0.523 | 0.301 | +0.01% | 0.060 | 0.476 |
| 4 | 214 | +0.00% | -0.030 | 0.488 | +0.07% | 0.666 | 0.253 |
| 5 | 214 | +0.01% | 0.064 | 0.474 | -0.13% | -1.209 | 0.114 |
| 6 | 214 | +0.00% | 0.030 | 0.488 | -0.08% | -0.716 | 0.237 |
| 7 | 214 | -0.02% | -0.238 | 0.406 | -0.09% | -0.850 | 0.198 |
| 8 | 214 | -0.08% | -0.717 | 0.237 | +0.02% | 0.171 | 0.432 |
| 9 | 214 | +0.06% | 0.520 | 0.302 | +0.11% | 0.880 | 0.190 |
| 10 | 214 | +0.08% | 0.761 | 0.224 | -0.03% | -0.237 | 0.407 |

Table 5 describes the daily (AAVs) and cumulative (CAAVs) average abnormal volumes observed over the [0, 10] trading day window following the release of green bond-related information. Baseline expected volumes are calculated as the average daily trading volume observed over the month before the release of the green bond-related information, in accordance with Jansen (2015). An immediate 11% increase in trading volume is observed, on average, on the day information is leaked about an upcoming green bond announcement (Table 5, Panel B). This indicates an immediate diffusion of information to the market that is acted upon by market traders, despite the muted price impact found in Table 4 (Panel B). It further highlights that Bloomberg News and First Word are economically significant sources of market information. Modest yet non-significant increases in trading volume are consistently observed over the trading week [1,5] following the release of information leaks, highlighting sustained investor interest. This contributes to an economically significant increase in the cumulative trading volume on the stock, with a cumulative rise of 61.3% in trading volumes, on average, in the 10 trading days following the information leaks.

Mixed results are found when considering trading dynamics following public announcements of green bonds. Investor interest is initially unchanged, before a weakly significant fall of 5% in trading volume the day after the public announcement.

These are followed by significant trading volume increases of 13-24% observed 2-7 trading days after public announcements. These contribute to a highly significant cumulative increase of 56% in trading volume over the 10 trading days after the public announcement of green bonds relative to the trading activity in the month before the announcement. In summary, the release of green bond-related information attracts heightened investor attention both after public announcements and news leaks, with the latter having a more consistent and positive impact on trading volumes.

**Table 5:** Daily (AAVs) and Cumulative (CAAVs) Average Abnormal Volumes observed over the 10 trading days following the release of green bond-related information. Abnormal volumes are calculated using the methodology proposed by Jansen (2015). *, **, and *** indicate significance at the 10%, 5%, and 1% level, respectively.

| Day from Announcement | n | Panel A. Public Announcement | | | | Panel B. Leaked News | | | |
|---|---|---|---|---|---|---|---|---|---|
| | | AAV$^{sig}$ | t-stat | CAAV$^{sig}$ | t-stat | AAV$^{sig}$ | t-stat | CAAV$^{sig}$ | t-stat |
| 0 | 214 | -0.01% | -0.002 | -0.01% | -0.002 | **+11.41%**** | 1.670 | **+11.41%**** | 1.667 |
| 1 | 214 | **-5.32%*** | -1.395 | -5.31% | -0.774 | +6.86% | 1.142 | **+18.19%**** | 1.772 |
| 2 | 214 | **+15.20%**** | 2.256 | +9.82% | 0.901 | +2.51% | 0.455 | **+20.69%*** | 1.500 |
| 3 | 214 | -2.79% | -0.623 | +7.04% | 0.562 | +6.75% | 1.242 | **+27.41%**** | 1.688 |
| 4 | 214 | -2.68% | -0.575 | +4.37% | 0.287 | +3.27% | 0.557 | **+30.67%*** | 1.636 |
| 5 | 214 | **+12.97%*** | 1.394 | +17.29% | 0.927 | +7.68% | 1.182 | **+38.32%**** | 1.764 |
| 6 | 214 | +0.61% | 0.080 | +17.90% | 0.817 | -1.10% | -0.253 | **+37.22%*** | 1.598 |
| 7 | 214 | **+24.05%**** | 1.681 | **+41.85%*** | 1.359 | +0.84% | 0.127 | **+38.05%*** | 1.536 |
| 8 | 214 | +7.52% | 1.234 | **+49.34%*** | 1.436 | -0.72% | -0.146 | **+37.34%*** | 1.400 |
| 9 | 214 | -0.64% | -0.152 | **+48.70%*** | 1.360 | **+15.32%*** | 1.534 | **+52.59%**** | 1.726 |
| 10 | 214 | **+7.77%*** | 1.301 | **+56.44%*** | 1.498 | **+8.74%*** | 1.447 | **+61.29%**** | 1.863 |

### *4.1.3. Wider Event Windows: Abnormal Returns*

We extend our abnormal return analysis to consider wider event windows surrounding public announcements and following information leaks. We also account for several market factors that could otherwise explain individual stock returns. The abnormal returns in Table 6 are the returns observed in excess of the expected returns predicted by the Fama-French 3-factor and Carhart 4-factor models. Persistent and negative returns are observed over the three days following the release of information related to green bonds, both following the release of pre-public news leaks (Table 6, Panel B) and around the time of the public announcement (Table 6, Panel A), with stock prices falling between 25 and 47 basis points, on average. These significant negative reactions intensify over the [0,10] trading day window, reaching a cumulative fall of 78 basis points 10 trading days after news leaks (Panel B). The same effect is observed more moderately following public announcements, with a 40 basis point drop 10 trading days after the release of information, on average (Panel A). There is a significant mid-term reversal in the extent to which stock prices fall, however, suggesting an initial overreaction to both leaked news and public announcements. Over the [11, 21] event window, stock prices rose between 72-86 basis points on average following both the public announcement of green bonds and the release of leaked information about upcoming green bond announcements. These mid-term increases in share prices are significant at the 5% level.

We also observe a significant fall of 50 to 76 basis points in share prices in the days before and after the public announcement of green bonds, when measured over [-3, 3] and [-5, 5] trading day event windows (Panel A). These results are not entirely unexpected. Howton et al. (1998) highlight that debt issues are not completely unanticipated by market investors and that investor reactions to bond announcements may not be fully reflective of the information within the announcements alone. Chaplinsky and Hansen (1993), for example, find negative stock returns for up to 140 days before the announcement of upcoming debt issuances. Given our sample is comprised of green bond announcements for which pre-public information was leaked on the Bloomberg Terminal news feed, it is not surprising that an increase in price volatility is observed pre- and post-announcement.

However, it does raise a question on the direction of the price impact. The [0, 3] trading day event window sees a fall of 33 to 38 basis points, intensifying to a fall of 67 to 76 basis points when viewed over the [-3, 3] event window. This implies that the [-3, -1] abnormal returns are also notably negative[13]. So why do arbitrageurs – or, in our sample, those with privileged information courtesy of their access to information leaks on Bloomberg – not try to benefit by buying before the public announcement and selling after (i.e. 'buying the rumour and selling the news')? In such a scenario, we expect share prices to increase in the days leading up to the announcement. Shleifer and Vishny (1997) reason arbitrage opportunities are not acted upon as commonly as expected, as arbitrage opportunities remain risky even when sufficient opportunities present. They argue that anomalies are enticing opportunities for arbitrageurs when the pattern of returns is not very noisy and the payoff horizon is short. While condition two may apply in our sample, condition one does not. There are numerous reasons why arbitrageurs may not be sufficiently incentivised to trade underlying equities around green bond announcements when news leakages occur. Chief among these would be the disparity of time between leaks, the heterogeneity in size, liquidity, and geographical scope of the equities in our 'leak' sample, and the market frictions associated with trading a global pool of equities.

---

[13] The abnormal returns on the day of the public announcement are -18 and -21 basis points according to the Fama-French and Carhart models, implying a cumulative abnormal return over the [-3, -1] event window of -34 and -38 basis points, respectively. Although not fully documented here, the same pattern can be observed over the [-5, 5] trading day event window.

**Table 6:** Cumulative Average Abnormal Returns (CAARs) observed over various event windows surrounding the public announcement of green bond issues (Panel A) and following the leakage of information relating to upcoming green bond announcements (Panel B). The abnormal returns are calculated according to the Fama-French 3-factor model and Carhart 4-factor model. *, **, and *** indicate significance at the 10%, 5%, and 1% level, respectively.

| Event Window | n | Panel A. Public Announcement | | Panel B. Leaked News | |
|---|---|---|---|---|---|
| | | **Fama-French** | **Carhart** | **Fama-French** | **Carhart** |
| | | CAAR$^{sig}$ ($t$-stat) | CAAR$^{sig}$ ($t$-stat) | CAAR$^{sig}$ ($t$-stat) | CAAR$^{sig}$ ($t$-stat) |
| [-10,10] | 214 | -0.38% (0.68) | -0.54% (-0.94) | – | – |
| [-5,5] | 214 | **-0.50%* (-1.36)** | **-0.62%* (-1.61)** | – | – |
| [-3,3] | 214 | **-0.67%*** (-2.54)** | **-0.76%*** (-2.75)** | – | – |
| [0,1] | 214 | **-0.26%** (-2.04)** | **-0.29%** (-2.09)** | **-0.21%* (-1.34)** | **-0.26%** (-1.68)** |
| [0,2] | 214 | **-0.25%* (-1.48)** | **-0.27%* (-1.58)** | **-0.38%** (-2.16)** | **-0.47%*** (-2.66)** |
| [0,3] | 214 | **-0.33%** (-1.73)** | **-0.38%** (-1.88)** | **-0.26%* (-1.34)** | **-0.33%** (-1.68)** |
| [0,5] | 214 | -0.27% (-1.09) | **-0.33%* (-1.29)** | **-0.36%* (-1.40)** | **-0.48%** (-1.80)** |
| [0,7] | 214 | -0.32% (-1.06) | **-0.41%* (-1.31)** | **-0.57%** (-1.92)** | **-0.73%** (-2.33)** |
| [0,10] | 214 | -0.28% (-0.78) | -0.40% (-1.05) | **-0.58%* (-1.43)** | **-0.78%** (-1.85)** |
| [11,21] | 214 | **+0.78%** (2.10)** | **+0.82%** (2.15)** | **+0.72%** (1.97)** | **+0.86%** (2.31)** |
| [21,30] | 214 | -0.28% (-0.82) | -0.33% (-0.96) | -0.17% (-0.52) | -0.22% (-0.66) |

### 4.2. Explaining Abnormal Returns after Information Leaks

In addition to determining the impact of information leaks on the trading activity of the firm's underlying equity, we explore the mechanisms by which these news leaks impact financial markets. Table 7 describes key drivers affecting investor reactions to corporate actions, as measured through abnormal returns. We again split our sample into financials (Table 7, Panel B) and non-financials (Table 7, Panel A). Of primary interest are the influence of bond characteristics on equity investor reactions. These include the issue's size, number of securities, average term to maturity and coupon rate, and embedded options within the issue structure. We also collect financial information relating to the firm, with proxy variables accounting for firm size, profitability, leverage, liquidity, and experience in the green bond market.

Table 7 highlights that abnormal returns are more predictable for non-financial equities than financial equities. Panel A shows that over the [0, 1] and [0, 2] windows following the release of leaked information, 29.3% and 23.5% of the variance in abnormal returns can be explained, respectively. Both model fits are statistically significant, and reveal much more information about investor reactions than the abnormal returns observed after public announcements (Appendix D). Generally, there is a positive baseline reaction to leaks about upcoming green bond announcements from non-financials, and investors respond positively to higher coupon rates and the presence of embedded options. However, these effect sizes are not significant. The most consistently significant drivers of abnormal returns for non-financials are Tobin's Q and free cash flow, which have a significantly positive effect on abnormal returns (Table 7, Panel A). Both factors have been highlighted as having a notable influence on equity market reactions to corporate actions around the issuance of debt. Chang et al. (2007) find that, when issuing secured debt, firms with higher Tobin's Q and higher free cash flows experience significantly higher abnormal returns than firms with lower Tobin's Q and lower free cash flows. These positive abnormal returns intensify when the firms have both high Tobin's Q and high free cash flow. In a green bond context, Baulkaran (2019) find investors react negatively to the announcement of green bond issues when the issuing firms have a higher cash flow ratio but react positively when they have a high Tobin's Q. The mixed findings on Tobin's Q and free cash flow can be reasoned. Tobin's Q is a proxy for a firm's investment opportunity set, with high-q firms seen as having greater investment opportunities (Howton et al. 1998). Furthermore, it can be argued that firms with relatively fewer growth opportunities (while maintaining strong governance structures) are more likely to have free cash flow (Amin et al. 2023). This could send a negative signal to the market when further debt is issued by firms with a diverse ownership group (Brush *et al.* 2000). However, free cash flows ultimately represent the flexibility for firms to invest in profit-maximising investment opportunities. We therefore posit the positive effects of free cash flow and Tobin's Q in our sample are indicative of the market's perception of the potential transformative impact of green financing. A combined high Tobin's Q and free cash flow represents the opportunity for firms to develop transformative sustainability-related business models, and the issuance of green bonds over conventional bonds signals their intention to do so.

These same dynamics do not apply to investor reactions to financial firms (Table 7, Panel B). There is a negative baseline reaction to the release of pre-public information, while higher Tobin's Q and higher free cash flows have a negative effect on abnormal returns. These effects are not significant. Instead, investors appear to value the operational efficiency of the issuer, with higher ROAs having a significant positive effect on abnormal returns. Positive effects are also found for the planned issue size and the issuing firm's debt-to-equity, while negative effects are found for longer-term structures and higher coupon rates. These effects are not significant, but they do paint a broad portrait of investor reactions. Taken collectively, these effect sizes may indicate that established, efficiently operating firms with the capacity to take on significant levels of debt are generally better received by the market, despite the negative overall reaction. From a green financing perspective, however, the results of Table 7, Panel B may reflect our earlier hypothesis that green bond leaks send a negative market signal when issued by financials due to the stringent regulations associated with the intended use of green bond proceeds. Regulatory requirements may hinder the capacity of financial institutions to develop innovative and transformative business models using green financing, acting as a negative signal when news of their impending announcement is leaked.

**Table 7:** Regression results describing Cumulative Abnormal Returns (CARs) observed in the two trading days following the release of pre-public information relating to an upcoming green bond announcement. We control for time-of-day, day-of-week, year, and region-specific effects. *, **, and *** indicate significance at the 10%, 5%, and 1% level, respectively.

| Dependent Variable: | Panel A: Non-financial Institutions | | | | | | Panel B: Financial Institutions | | | | | |
|---|---|---|---|---|---|---|---|---|---|---|---|---|
| | CAR [0, 1] | | | CAR [0, 2] | | | CAR [0, 1] | | | CAR [0, 2] | | |
| CARs related to Leaks | (1) | (2) | (3) | (4) | (5) | (6) | (7) | (8) | (9) | (10) | (11) | (12) |
| Issue Size | -0.194 (0.351) | -0.269 (0.346) | -0.333 (0.356) | -0.113 (0.399) | -0.209 (0.391) | -0.289 (0.402) | – | – | – | – | – | – |
| Term | -0.052 (0.042) | -0.036 (0.042) | -0.036 (0.043) | -0.067 (0.048) | -0.047 (0.048) | -0.048 (0.048) | – | – | – | – | – | – |
| Cpn | 0.086 (0.136) | 0.117 (0.134) | 0.104 (0.137) | 0.121 (0.154) | 0.161 (0.151) | 0.149 (0.155) | – | – | – | – | – | – |
| Option in Issue | **1.413* (0.730)** | 1.154 (0.728) | 1.152 (0.761) | 1.276 (0.830) | 0.945 (0.821) | 0.886 (0.857) | – | – | – | – | – | – |
| Bonds in Issue | 0.173 (0.461) | 0.304 (0.457) | 0.380 (0.470) | -0.397 (0.525) | -0.228 (0.516) | -0.140 (0.530) | – | – | – | – | – | – |
| FTI | 0.897 (0.625) | 0.605 (0.630) | 0.568 (0.636) | 0.720 (0.711) | 0.347 (0.711) | 0.299 (0.716) | – | – | – | – | – | – |
| ROA | -7.699 (9.217) | -10.222 (9.129) | -11.554 (9.42) | 3.992 (10.477) | 0.767 (10.307) | -1.377 (10.614) | – | – | – | – | – | – |
| D/E | -0.150 (0.205) | -0.175 (0.201) | -0.236 (0.216) | 0.006 (0.232) | -0.026 (0.227) | -0.115 (0.244) | – | – | – | – | – | – |
| FCF | 0.070 (0.054) | **0.098* (0.055)** | **0.102* (0.057)** | 0.101 (0.061) | **0.136** (0.062)** | **0.138** (0.064)** | – | – | – | – | – | – |
| Tobin's Q | **0.670** (0.290)** | **0.800*** (0.291)** | **0.830*** (0.296)** | 0.372 (0.329) | 0.538 (0.329) | **0.571* (0.334)** | – | – | – | – | – | – |
| Issue Size | – | – | – | – | – | – | 0.189 (0.227) | 0.175 (0.232) | 0.153 (0.241) | 0.122 (0.269) | 0.045 (0.270) | 0.018 (0.281) |
| Term | – | – | – | – | – | – | -0.011 (0.107) | -0.012 (0.107) | -0.017 (0.110) | -0.010 (0.126) | -0.016 (0.125) | -0.020 (0.128) |
| Cpn | – | – | – | – | – | – | -0.112 (0.161) | -0.104 (0.163) | -0.135 (0.170) | -0.046 (0.191) | -0.002 (0.191) | -0.030 (0.198) |
| Option in Issue | – | – | – | – | – | – | 0.207 (0.598) | 0.241 (0.608) | 0.340 (0.622) | -0.093 (0.708) | 0.100 (0.709) | 0.197 (0.726) |
| Bonds in Issue | – | – | – | – | – | – | 0.252 (0.423) | 0.246 (0.425) | 0.250 (0.429) | -0.113 (0.500) | -0.149 (0.496) | -0.142 (0.501) |
| FTI | – | – | – | – | – | – | 0.122 (0.483) | 0.110 (0.486) | 0.111 (0.492) | -0.135 (0.572) | -0.198 (0.567) | -0.193 (0.574) |
| ROA | – | – | – | – | – | – | **18.260* (10.248)** | **17.726* (10.398)** | **19.235* (11.506)** | 17.884 (12.127) | 14.887 (12.130) | 16.747 (13.437) |
| D/E | – | – | – | – | – | – | 0.056 (0.113) | 0.057 (0.113) | 0.085 (0.119) | 0.002 (0.134) | 0.005 (0.132) | 0.032 (0.139) |
| FCF | – | – | – | – | – | – | -0.026 (0.035) | -0.027 (0.035) | -0.029 (0.036) | -0.033 (0.042) | -0.036 (0.041) | -0.038 (0.042) |
| Tobin's Q | – | – | – | – | – | – | -0.331 (1.002) | -0.286 (1.014) | -0.117 (1.332) | -0.869 (1.186) | -0.615 (1.183) | -0.513 (1.555) |
| Constant | 1.115 (6.567) | 0.550 (6.451) | 1.816 (6.670) | 0.200 (7.465) | -0.522 (7.284) | 1.015 (7.515) | -4.385 (4.286) | -4.310 (4.310) | -3.449 (4.573) | -2.623 (5.072) | -2.199 (5.028) | -1.255 (5.340) |
| Time of Day | ✓ | ✓ | ✓ | ✓ | ✓ | ✓ | ✓ | ✓ | ✓ | ✓ | ✓ | ✓ |
| Day | ✓ | ✓ | ✓ | ✓ | ✓ | ✓ | ✓ | ✓ | ✓ | ✓ | ✓ | ✓ |
| Year | | ✓ | ✓ | | ✓ | ✓ | | ✓ | ✓ | | ✓ | ✓ |
| Region | | | ✓ | | | ✓ | | | ✓ | | | ✓ |
| Observations | 100 | 100 | 100 | 100 | 100 | 100 | 114 | 114 | 114 | 114 | 114 | 114 |
| $R^2$ | 25.0% | 28.6% | 29.3% | 17.4% | 22.4% | 23.5% | 14.3% | 14.4% | 15.1% | 11.2% | 13.8% | 14.3% |
| F-stat | 2.210** | 2.435*** | 2.150** | 1.398 | 1.754* | 1.591* | 1.284 | 1.191 | 1.081 | 0.966 | 1.130 | 1.012 |

# 5  Concluding Remarks

Baulkaran (2019) and Tang and Zhang (2020) raise the possibility of information leakages related to upcoming green bond issues. Our research study is the first to confirm this effect through the presence of information leaks on the Bloomberg Terminal. We find evidence these leaks have a significant impact on the trading dynamics of underlying firms, both on an intraday and daily level. The price and trading impact is immediate and persists over the days following the release of information. Investors react negatively to both the pre-public leakage and public announcement of upcoming green bond issues, with the impact being stronger following the leaks. When viewed over a longer event window (11-20 trading days after information releases), we find moderate evidence of market overreaction. The analysis of abnormal trading volumes also reveals significant increases following both leaks and announcements, with trading volumes spiking more prominently and consistently after leaks.

Our investigations into the price impact of individual leaks reveal several dynamics that align with common conventions on investor reactions to corporate actions and news releases. When considering the impact of the timing of the leaks, we find that intraday news released at least 1 hour into the trading day is associated with a significant increase in abnormal returns and a significant reduction in trading volumes. In contrast, pre- and early-market news (including information leaked following the previous day's close) is associated with a significant fall in abnormal returns and a rise in trading volumes. This is consistent with Patell and Wolfson's (1982) findings that 'good news' is released during trading hours, while 'bad news' is withheld until after market close. Furthermore, investors tend to react positively to green bond leaks associated with non-financial firms in the minutes following the leak, before reverting to a mixed intraday reaction. In contrast, a persistent negative intraday reaction is seen for news leaks related to financial institutions. Exploring the underlying mechanisms driving these differences in abnormal returns, we find investors positively react to non-financials with higher levels of free cash flows and Tobin's Q, and when there is an option embedded within the planned bond structure. These findings agree with those of Chang *et al.* (2007). For financial firms, investors are positively receptive to firms with higher levels of ROA. When considered collectively, the firm- and bond-level variables we use to explain abnormal returns paint two broad pictures of investor reactions to financials and non-financials. Investors appear to be more receptive to non-financials announcing green bonds when the firms are well-positioned to develop transformative sustainability-related business models. In contrast, investors appear to be more pessimistic about the potentially transformative impact of green financing for financial institutions, instead preferring when established, efficiently operating firms with the capacity to take on significant levels of debt plan green bond issues.

The significant market impact that leaked news has in our sample raises broader questions on informational inequity and informational sourcing strategies for event studies. Information leakages have re-emerged as a prominent topic in financial literature due to the increasing number of channels through which financially material information can be disseminated through news media (Bernile *et al.* 2016; Cieslak *et al.* 2019). News media has a significant and causal influence on trading activities (Peress 2014), with Fedyk (2024) highlighting that the Bloomberg News screen itself plays a significant role in influencing institutional investor actions. The influence of the Bloomberg News screen is reflected in our results.

The proliferation of information sources for financially material data has wider implications for the validity of findings from event studies that rely on data drawn from financial databases. For example, the leaks we found in our research studies account for 7.5% of all green bond announcements made by publicly traded firms over the study period, a non-negligible figure (Appendix C). The extent of the pricing activity that occurred based on the release of this pre-public information suggests that green bond event studies relying on announcement date data from the Bloomberg SRCH database may need to be revisited. However, establishing the specific diffusive effect of leaked news is difficult. This is due in part to inherent limitations on the ability to source news leakages (Bernile *et al.* 2016), and due in part to the fragmentation of access to news sources reporting financially-material data (Fedyk 2024), leading to pricing inefficiencies. Although our sample of 259 leaked news items are the earliest records found by the authorship team on information being released about upcoming green bond announcements, it is plausible this information was released and acted upon even earlier. Institutional investors make more informed decisions before the announcement of financially material news (Ke and Petroni 2004), and given Bloomberg is predominantly used by sophisticated institutional investors, it is possible anticipatory trading was carried out in advance of the leaks. Indeed, we find evidence of this courtesy of the significantly negative abnormal returns observed in the [-3, -1] event window leading up to the leak, so the presence of pre-leakage leakages cannot be ruled out. Nevertheless, our research addresses a gap in understanding how pre-public information can affect market reactions before official public announcements and provides greater clarity on market reactions to, and perceptions of, green bond announcements.

## Declaration of Competing Interests

The authors declare no competing interests as part of this research.

## Funding statement

The authors acknowledge the support of their university department while declaring that no external funding was received to conduct the research in this study.

# Appendix A

**Table A1:** Description of variables used in our regression analysis to further explain abnormal returns, along with the sources of the data.

| Variables | Definition | Source |
|---|---|---|
| **Panel A: Green Bond Characteristics** | | |
| Issue Size | = natural log of the total amount issued | Bloomberg |
| Term | = average time to maturity of bonds in issue (years) | Bloomberg |
| Cpn | = average coupon rate of bonds in issue (%) | Bloomberg |
| Option in Issue | = presence of embedded option in bond issue (0/1) | Bloomberg |
| Bonds in Issue | = number of bonds in bond issue | Bloomberg |
| **Panel B: Firm-level Characteristics** | | |
| FTI | = whether the firm previously issued a Green Bond (0/1) | Bloomberg |
| MktCap | = natural log of 3-year averaged market capitalisation | S&P Capital IQ Pro |
| Assets | = natural log of 3-year averaged total assets | S&P Capital IQ Pro |
| ROA | = 3-year averaged return on assets (%) | S&P Capital IQ Pro |
| D/E | = 3-year averaged total debt / common equity (x) | S&P Capital IQ Pro |
| FCF | = natural log of 3-year averaged free cash flow | S&P Capital IQ Pro |
| Tobin's Q | = avg. 3Y Market Cap / avg. 3Y Total Assets (x) | S&P Capital IQ Pro |
| **Panel C: Outcome Variables** | | |
| CAR[0,1] | Cumulative Abnormal Return over [0,1] window after news leak | CRSP |
| CAR[0,2] | Cumulative Abnormal Return over [0,2] window after news leak | CRSP |

# Appendix B

**Table B1:** Correlations between the variables described in Appendix A, which are used in our regression analysis to further explain abnormal returns.

| | Issue Size | Term | Cpn | Option in Issue | Bonds in Issue | FTI | MktCap | Assets | ROA | D/E | FCF | Tobin's Q |
|---|---|---|---|---|---|---|---|---|---|---|---|---|
| Issue Size | 1 | – | – | – | – | – | – | – | – | – | – | – |
| Term | 0.14 | 1 | – | – | – | – | – | – | – | – | – | – |
| Cpn | 0.04 | 0.14 | 1 | – | – | – | – | – | – | – | – | – |
| Option in Issue | 0.34 | 0.36 | 0.23 | 1 | – | – | – | – | – | – | – | – |
| Bonds in Issue | 0.33 | -0.09 | 0.14 | -0.09 | 1 | – | – | – | – | – | – | – |
| FTI | 0.01 | 0.09 | 0.08 | 0.21 | -0.24 | 1 | – | – | – | – | – | – |
| MktCap | 0.31 | <.01 | -0.09 | -0.01 | 0.12 | -0.23 | 1 | – | – | – | – | – |
| Assets | 0.16 | -0.13 | -0.11 | -0.19 | 0.10 | -0.26 | 0.66 | 1 | – | – | – | – |
| ROA | -0.27 | -0.05 | -0.02 | 0.02 | -0.05 | 0.08 | -0.07 | -0.29 | 1 | – | – | – |
| D/E | 0.11 | -0.09 | -0.03 | -0.17 | -0.04 | -0.16 | 0.08 | 0.33 | -0.42 | 1 | – | – |
| FCF | 0.24 | 0.01 | -0.04 | 0.08 | 0.04 | 0.13 | 0.35 | 0.28 | -0.11 | 0.08 | 1 | – |
| Tobin's Q | -0.13 | 0.07 | 0.25 | 0.17 | -0.02 | 0.16 | -0.06 | -0.25 | 0.47 | -0.28 | -0.07 | 1 |

# Appendix C

**Table C1:** Distribution of 238 information leakage events stemming from 170 publicly traded, highly liquid stocks from a sample of 60 exchanges. Also included is the availability of factors to explain stock returns in each market, which are used to isolate the significance of the news leaks when examining abnormal returns.

| Exchange | Country | Green Bonds Announced | Leaks | Meeting Liquidity Condition | Factor Coverage |
|---|---|---|---|---|---|
| Abu Dhabi Securities Exchange | United Arab Emirates | 15 | 0 | 0 | Not Covered |
| Athens Stock Exchange | Greece | 9 | 1 | 1 | Accurate to Region |
| Australian Securities Exchange | Australia | 35 | 3 | 3 | Accurate to Region |
| Bolsa de Comercio de Santiago de Chile | Chile | 18 | 1 | 1 | Not Covered |
| Bolsa de Valores de Colombia | Colombia | 7 | 0 | 0 | Not Covered |
| Bolsa de Valores de Lima S.A. | Peru | 1 | 0 | 0 | Not Covered |
| Bolsas y Mercados Argentinos Exchange | Argentina | 2 | 0 | 0 | Not Covered |
| Borsa Istanbul | Turkey | 12 | 1 | 1 | Not Covered |
| Borsa Italiana | Italy | 60 | 14 | 14 | Accurate to Region |
| Botswana Stock Exchange | Botswana | 1 | 0 | 0 | Not Covered |
| Bratislava Stock Exchange | Hungary | 1 | 0 | 0 | Not Covered |
| Budapest Stock Exchange | Hungary | 7 | 1 | 1 | Not Covered |
| Bursa Malaysia | Malaysia | 29 | 0 | 0 | Not Covered |
| Dow Jones Industrial Average | United States | 14 | 0 | 0 | Not Covered |
| Euronext Amsterdam | Netherlands | 43 | 7 | 7 | Accurate to Region |
| Euronext Brussels | Belgium | 17 | 1 | 1 | Accurate to Region |
| Euronext Dublin | Ireland | 12 | 0 | 0 | Not Covered |
| Euronext Lisbon | Portugal | 18 | 0 | 0 | Not Covered |
| Euronext Paris | France | 408 | 13 | 13 | Accurate to Region |
| First North Stockholm | Sweden | 95 | 0 | 0 | Not Covered |
| FTSE All-Share | United Kingdom | 66 | 8 | 6 | Accurate to Region |
| Hong Kong Stock Exchange | Hong Kong | 302 | 14 | 14 | Accurate to Region |
| IBEX 35 | Spain | 132 | 2 | 2 | Accurate to Region |
| IBEX Medium Cap | Spain | 1 | 1 | 1 | Not Covered |
| IBEX Small Cap | Spain | 1 | 1 | 1 | Not Covered |
| Indonesia Stock Exchange | Indonesia | 18 | 0 | 0 | Not Covered |
| Johannesburg Stock Exchange | South Africa | 21 | 2 | 2 | Not Covered |
| KOSPI Stock Market | South Korea | 246 | 13 | 13 | Not Covered |
| Mexico Stock Exchange | Mexico | 4 | 0 | 0 | Not Covered |
| Namibian Stock Exchange | Namibia | 3 | 0 | 0 | Not Covered |
| Nasdaq Baltic - Vilnius | Lithuania | 1 | 0 | 0 | Not Covered |
| Nasdaq Capital Market | United States | 1 | 0 | 0 | Exact |
| Nasdaq Global Market | United States | 4 | 0 | 0 | Exact |
| Nasdaq Global Select | United States | 209 | 2 | 2 | Exact |
| NASDAQ OMX Copenhagen | Denmark | 37 | 4 | 4 | Accurate to Region |
| NASDAQ OMX Helsinki | Finland | 23 | 7 | 6 | Accurate to Region |
| NASDAQ OMX Iceland | Iceland | 4 | 0 | 0 | Not Covered |
| NASDAQ OMX Stockholm | Sweden | 134 | 50 | 46 | Accurate to Region |
| National Stock Exchange of India | India | 26 | 9 | 9 | Accurate to Region |
| New York Stock Exchange | United States | 197 | 15 | 13 | Exact |
| New Zealand Exchange | New Zealand | 13 | 3 | 2 | Accurate to Region |
| Nigerian Exchange Limited | Nigeria | 1 | 0 | 0 | Not Covered |
| Oslo Bors | Norway | 149 | 14 | 14 | Accurate to Region |
| OTC-X Berner KantonalBank | Switzerland | 31 | 0 | 0 | Not Covered |
| Philippine Stock Exchange | Philippines | 14 | 2 | 2 | Not Covered |
| Qatar Exchange | Qatar | 1 | 0 | 0 | Not Covered |
| Sao Paulo Stock Exchange | Brazil | 24 | 1 | 1 | Not Covered |
| Saudi Arabian Stock Exchange | Saudi Arabia | 3 | 0 | 0 | Not Covered |
| Shanghai Stock Exchange | China | 219 | 6 | 5 | Exact |
| Shenzhen Stock Exchange | China | 17 | 2 | 2 | Accurate to Region |
| Singapore Exchange | Singapore | 17 | 1 | 1 | Accurate to Region |
| SIX Swiss Exchange | Switzerland | 10 | 9 | 6 | Accurate to Region |
| Stock Exchange of Thailand | Thailand | 35 | 0 | 0 | Not Covered |
| Taiwan Stock Exchange | Taiwan | 41 | 0 | 0 | Not Covered |
| Tel Aviv Stock Exchange | Israel | 2 | 1 | 1 | Not Covered |
| Tokyo Stock Exchange | Japan | 295 | 26 | 21 | Exact |
| TSX Toronto Exchange | Canada | 49 | 1 | 1 | Exact |
| Vienna Stock Exchange | Austria | 64 | 7 | 6 | Accurate to Region |
| Warsaw Stock Exchange | Poland | 13 | 1 | 0 | Not Covered |
| XETRA | Germany | 230 | 15 | 15 | Accurate to Region |
| **Sum** | | **3,462** | **259** | **238** | |

# Appendix D

**Table D1:** Regression results describing Cumulative Abnormal Returns (CARs) observed in the two trading days following the public announcement of the green bond (mimicking the observational period of the News Leaks in Table 7). We control for day-of-week, year, and region-specific effects. Returns are less explainable over these trading windows than the corresponding trading window following information leakages. *, **, and *** indicate significance at the 10%, 5%, and 1% level, respectively.

| Dependent Variable: | Panel A. Non-financial Institutions | | | | | | Panel B. Financial Institutions | | | | | |
|---|---|---|---|---|---|---|---|---|---|---|---|---|
| | CAR [0, 1] | | | CAR [0, 2] | | | CAR [0, 1] | | | CAR [0, 2] | | |
| CARs related to PAs | (1) | (2) | (3) | (4) | (5) | (6) | (7) | (8) | (9) | (10) | (11) | (12) |
| Issue Size | 0.025 (0.268) | 0.036 (0.270) | 0.001 (0.278) | -0.079 (0.313) | -0.071 (0.315) | -0.125 (0.326) | – | – | – | – | – | – |
| Term | -0.012 (0.032) | -0.015 (0.033) | -0.018 (0.033) | -0.062 (0.038) | **-0.064 (0.039)*** | **-0.066 (0.039)*** | – | – | – | – | – | – |
| Cpn | 0.107 (0.104) | 0.101 (0.105) | 0.106 (0.107) | -0.077 (0.121) | -0.081 (0.123) | -0.085 (0.125) | – | – | – | – | – | – |
| Option in Issue | -0.121 (0.558) | -0.065 (0.570) | -0.206 (0.595) | 0.190 (0.650) | 0.227 (0.665) | 0.151 (0.696) | – | – | – | – | – | – |
| Bonds in Issue | -0.325 (0.351) | -0.345 (0.355) | -0.319 (0.367) | -0.348 (0.410) | -0.361 (0.414) | -0.303 (0.429) | – | – | – | – | – | – |
| FTI | 0.690 (0.476) | 0.752 (0.493) | 0.725 (0.495) | 0.823 (0.555) | 0.865 (0.575) | 0.836 (0.580) | – | – | – | – | – | – |
| ROA | 3.857 (6.952) | 4.182 (7.010) | 2.320 (7.230) | 1.703 (8.104) | 1.920 (8.181) | 0.045 (8.463) | – | – | – | – | – | – |
| D/E | 0.043 (0.153) | 0.043 (0.154) | -0.021 (0.167) | -0.067 (0.179) | -0.066 (0.180) | -0.136 (0.195) | – | – | – | – | – | – |
| FCF | **0.076 (0.041)*** | 0.071 (0.043) | 0.065 (0.044) | 0.031 (0.048) | 0.027 (0.050) | 0.026 (0.052) | – | – | – | – | – | – |
| Tobin's Q | -0.230 (0.222) | -0.256 (0.228) | -0.250 (0.231) | -0.344 (0.259) | -0.361 (0.267) | -0.343 (0.271) | – | – | – | – | – | – |
| Issue Size | – | – | – | – | – | – | -0.005 (0.182) | -0.001 (0.186) | -0.046 (0.193) | 0.157 (0.271) | 0.171 (0.276) | 0.099 (0.285) |
| Term | – | – | – | – | – | – | 0.036 (0.085) | 0.037 (0.086) | 0.041 (0.087) | -0.093 (0.127) | -0.092 (0.128) | -0.088 (0.129) |
| Cpn | – | – | – | – | – | – | **0.314 (0.125)**** | **0.312 (0.126)**** | **0.270 (0.133)**** | **0.465 (0.185)**** | **0.457 (0.188)**** | **0.371 (0.197)*** |
| Option in Issue | – | – | – | – | – | – | -0.795 (0.479)* | -0.805 (0.488) | -0.696 (0.497) | -0.208 (0.713) | -0.244 (0.725) | -0.031 (0.735) |
| Bonds in Issue | – | – | – | – | – | – | -0.370 (0.339) | -0.368 (0.341) | -0.353 (0.343) | -0.326 (0.505) | -0.319 (0.507) | -0.301 (0.507) |
| FTI | – | – | – | – | – | – | **-0.668 (0.384)*** | **-0.665 (0.387)*** | -0.622 (0.390) | -0.538 (0.571) | -0.527 (0.575) | -0.458 (0.577) |
| ROA | – | – | – | – | – | – | -2.825 (7.979) | -2.670 (8.103) | 1.247 (9.153) | -2.127 (11.868) | -1.584 (12.048) | 4.695 (13.54) |
| D/E | – | – | – | – | – | – | -0.111 (0.089) | -0.111 (0.090) | -0.076 (0.095) | -0.158 (0.133) | -0.159 (0.133) | -0.090 (0.140) |
| FCF | – | – | – | – | – | – | -0.009 (0.028) | -0.009 (0.028) | -0.011 (0.029) | -0.012 (0.042) | -0.012 (0.042) | -0.015 (0.042) |
| Tobin's Q | – | – | – | – | – | – | -0.119 (0.789) | -0.133 (0.799) | -0.284 (1.059) | 0.027 (1.173) | -0.019 (1.188) | -0.088 (1.566) |
| Constant | -1.380 (5.035) | -1.257 (5.062) | -0.687 (5.209) | 1.844 (5.870) | 1.926 (5.908) | 2.871 (6.096) | 0.336 (3.440) | 0.313 (3.461) | 1.507 (3.629) | -2.606 (5.117) | -2.684 (5.146) | -0.669 (5.368) |
| Time of Day | | | | | | | | | | | | |
| Day of Week | ✓ | ✓ | ✓ | ✓ | ✓ | ✓ | ✓ | ✓ | ✓ | ✓ | ✓ | ✓ |
| Year | | ✓ | ✓ | | ✓ | ✓ | | ✓ | ✓ | | ✓ | ✓ |
| Region | | | ✓ | | | ✓ | | | ✓ | | | ✓ |
| Observations | 100 | 100 | 100 | 100 | 100 | 100 | 114 | 114 | 114 | 114 | 114 | 114 |
| R:$^2$ | 11.5% | 11.8% | 13.2% | 15.6% | 15.7% | 16.6% | 10.7% | 10.7% | 12.3% | 8.9% | 9.0% | 11.5% |
| F-stat | 0.941 | 0.882 | 0.853 | 1.339 | 1.229 | 1.113 | 1.005 | 0.920 | 0.913 | 0.824 | 0.698 | 0.850 |